\documentstyle[prl,twocolumn,aps,graphicx]{revtex}
\begin{document}
\bibliographystyle{prsty}
\draft

 \newcommand{\mytitle}[1]{
 \twocolumn[\hsize\textwidth\columnwidth\hsize
 \csname@twocolumnfalse\endcsname #1 \vspace{1mm}]}

\mytitle{
\title{Theory of Diluted Magnetic Semiconductor Ferromagnetism}
\author{J\"urgen K\"onig$^1$, Hsiu-Hau Lin$^{1,2}$, 
        and Allan H. MacDonald$^1$}
\address{$^1$Department of Physics, Indiana University, Bloomington, IN 47405\\
        $^2$Department of Physics, National Tsing-Hua University, Hsinchu 300, 
        Taiwan}
\maketitle

\begin{abstract}

We present a theory of carrier-induced ferromagnetism in diluted magnetic 
semiconductors (${\rm III}_{1-x}{\rm Mn}_{x}{\rm V}$) which allows for 
arbitrary itinerant-carrier spin polarization and dynamic correlations.
Both ingredients are essential in identifying the system's elementary 
excitations and describing their properties.
We find a branch of collective modes, in addition to the spin waves and 
Stoner continuum which occur in metallic ferromagnets,
and predict that the low-temperature spin stiffness is independent of the 
strength of the exchange coupling between magnetic ions and itinerant carriers.
We discuss the temperature dependence of the magnetization and the heat 
capacity.

\end{abstract}
\pacs{75.30.Ds,75.40.Gb,75.50.Dd}
}

{\it Introduction.}
The recent discovery of carrier-induced ferromagnetism\cite{Story86,Ohno92/96}
in diluted magnetic semiconductors (DMS) has generated intense interest, in 
part because it opens the prospect of developing devices which combine 
information processing and storage functionalities in one material
\cite{Ohno99,Prinz98,Hayashi98,Pekarek98,VanEsch97,Oiwa99,Beschoten99}.
Critical temperatures $T_c$ exceeding $100 {\rm K}$ have been 
realized\cite{Ohno99} by using low-temperature molecular beam epitaxy growth 
to introduce a high 
concentration $c$ of randomly distributed ${\rm Mn}^{2+}$ ions in GaAs systems
with a high hole density $c^*$.
The tendency toward ferromagnetism and trends in the observed $T_c$'s have 
been explained using a 
picture\cite{Dietl97,Takahashi97,Jungwirth99,Dietl00,Lee00} in which uniform 
itinerant-carrier spin polarization mediates a long-range ferromagnetic
interaction between the ${\rm Mn}^{2+}$ ions with spin $S=5/2$.    

We present here the first theory which accounts for dynamic correlations in 
the ordered state and is able to describe its fundamental properties.
We use a path-integral formulation and, in the RKKY spirit of the previous 
theory, integrate out the itinerant carriers, and expand the effective action 
for the impurity spins up to quadratic order.
In addition to the usual Goldstone spin waves, we find itinerant-carrier 
dominated collective excitations analogous to the optical spin-waves in a 
ferrimagnet.
We also find that the spin stiffness is inversely proportional to the 
itinerant-carrier mass and independent of the strength of the exchange
coupling between magnetic ions and itinerant carriers, in contrast to the 
mean-field critical temperature which is proportional to the itinerant-carrier 
mass and to the square of the exchange coupling.  
We extend our calculations to the critical temperature using an approximate 
self-consistent spin-wave scheme and address the temperature-dependent
magnetization and specific heat.

{\it Hamiltonian and effective action.}
We study a model which provides an accurate\cite{modelrefs} description of Mn 
based zincblende DMS's.
Magnetic ions with $S=5/2$ at positions $\vec R_I$ are antiferromagnetically 
coupled to valence-band carriers described by envelope functions,
\begin{equation}
  H = H_0 + J_{pd} \int \hspace{-1mm}d^3 r \;
  \vec S(\vec r) \cdot \vec s(\vec r) ,
\end{equation}
where $\vec S(\vec r) = \sum_I \vec S_I \delta( \vec r - \vec R_I)$ is the 
impurity-spin density. 
The itinerant-carrier spin density  is expressed in terms of 
carrier field operators by 
$\vec s(\vec r)= {1\over 2} \sum_{\sigma \sigma'} \Psi^\dagger_\sigma (\vec r)
 \vec \tau_{\sigma \sigma'} \Psi_{\sigma'} (\vec r)$.
($\vec \tau$ is the vector of Pauli spin matrices.) 
$H_0$ includes the valence-band envelope-function Hamiltonian\cite{band}
and, if an external magnetic field $\vec B$ is present, the Zeeman energy,
\begin{eqnarray}
  H_0 &=& \int \hspace{-1mm}d^3 r\; \bigg\{ \sum_\sigma
      \hat\Psi^\dagger_\sigma (\vec r)
      \left( -{\hbar^2 \vec\nabla^2\over 2m^*}-\mu^{*} \right)
      \hat \Psi_\sigma (\vec r)
\nonumber \\
  &&- g\mu_B \vec B \cdot \vec S(\vec r)
 - g^* \mu_B \vec B \cdot \vec s (\vec r)\bigg\} .
\end{eqnarray}
The effective mass, chemical potential, and $g$-factor of the itinerant 
carriers are labeled by $m^*$, $\mu^*$, and $g^*$.
The model we use here is, thus, related to colossol magnetoresistance (CMR) 
materials\cite{Millis95} and 
identical to those for dense Kondo systems, which simplify when the 
itinerant-carrier density $c^*$ is much smaller than the magnetic ion density 
$c$\cite{Sigrist91}.
The fact that $c^*/c \ll 1$ in ferromagnetic-semiconductor materials is 
essential to their ferromagnetism.
Similar models have been used for ferromagnetism induced by magnetic 
ions in nearly ferromagnetic metals such as palladium\cite{Doniach67}.

We represent the impurity spins in terms of 
Holstein-Primakoff (HP) bosons\cite{Auerbach94}.
By coarse graining, the spin density 
$\vec S(\vec r)$ can be replaced by a smooth function,
\begin{eqnarray}
   S^+(\vec r) &=& \left(\sqrt{2cS - b^{\dag}(\vec r) b(\vec r)} \, \right)
   b(\vec r)
\\
   S^-(\vec r) &=& b^\dag(\vec r) \sqrt{2cS - b^{\dag}(\vec r) b(\vec r)}
\\
   S^z(\vec r) &=& cS - b^{\dag}(\vec r) b(\vec r)
\end{eqnarray}
with bosonic fields $b^{\dag}(\vec r), b(\vec r)$.
The partition function as a coherent-state path-integral in imaginary times
reads,
\begin{equation}
\label{partition function}
 Z = \int \hspace{-1mm}
 {\cal D} [\bar z z] {\cal D} [\bar\Psi \Psi]
 e^{-\int_0^\beta d\tau L( \bar z z, \bar\Psi \Psi)}
\end{equation}
with $L = \int d^3 r \left[ \bar z \partial_\tau z +
\sum_\sigma \bar \Psi_\sigma \partial_\tau \Psi_\sigma \right] +
H (\bar z z, \bar\Psi \Psi)$.
The bosonic (impurity spins) and fermionic (itinerant carriers) degrees of 
freedom are labeled by the complex variables $\bar z,z$ and the Grassmann 
numbers $\bar \Psi, \Psi$, respectively.

Since the Hamiltonian is bilinear in fermionic fields, we can integrate out
the itinerant carriers and arrive at an effective description in terms of the 
localized spin density only, $Z = \int {\cal D} [\bar z z] 
\exp (- S_{\rm eff} [\bar z z])$ with the action
\begin{eqnarray}
  S_{\rm eff} [\bar z z] = \int_0^\beta \hspace{-1mm}d\tau
  \int \hspace{-1mm}d^3 r \left[
  \bar z \partial_\tau z - g\mu_B B (cS - \bar zz)
  \right]
\nonumber \\
  - \ln \det \left[ (G^{MF})^{-1} + \delta G^{-1}(\bar zz) \right] \, .
\label{effective action}
\end{eqnarray}
Here, we have already split the total kernel $G^{-1}$ into a mean-field part 
$(G^{MF})^{-1}$ and a fluctuating part $\delta G^{-1}$,
\begin{eqnarray}
   (G^{MF})^{-1} &=&
 \left( \partial_\tau - {\hbar^2 \vec \nabla^2 \over 2m^*} -\mu^* \right)
 {\bf 1} + {\Delta \over 2}\tau^z
\\
   \delta G^{-1} &=& {J_{pd}\over 2} \left[
    (z \tau^- + \bar z \tau^+) \sqrt{2cS - \bar z z}  - \bar z z \tau^z
\right]
\end{eqnarray}
where $\Delta = cJ_{pd}S - g^*\mu_B B$ is the zero-temperature 
spin-splitting gap for the itinerant carriers.
The physics of the itinerant carriers is embedded in the effective action
of the magnetic ions.
It is responsible for the retarded and non-local character of the interactions 
between magnetic ions, which is described here for the first time.

{\it Independent spin-wave theory.}
The independent spin-wave theory, which is a good approximation at low 
temperatures, is obtained by expanding Eq.~(\ref{effective action}) up to 
quadratic order in $z$.
We find (in Fourier representation)
\begin{equation}
  S_{\rm eff} [\bar z z] = {1\over \beta V} \hspace{-1mm}
  \sum_{|\vec p| < p_c, m} \! \!
  \bar z(\vec p, \nu_m) D^{-1}(\vec p, \nu_m) z(\vec p, \nu_m),
\end{equation}
in addition to the temperature-dependent mean-field contribution.
A Debye cutoff ($p_c^3 = 6 \pi^2 c$) ensures that we include the correct 
number of magnetic ion degrees of freedom.
The kernel of the quadratic action is the inverse of the spin-wave propagator,
\begin{eqnarray}
   D^{-1}(\vec p, \nu_m) = 
 -i \nu_m + g\mu_B B + J_{pd} n^{*}
 \hspace*{2.cm}
\nonumber \\
 + { c J_{pd}^2 S \over 2\beta V} \sum_{n, \vec k}
 G^{MF}_\uparrow (\vec k, \omega_n)
 G^{MF}_\downarrow (\vec k+ \vec p, \omega_n+\nu_m)
\label{inverse propagator}
\end{eqnarray}
with the mean-field itinerant carrier Green's function
$G^{MF}_\sigma (\vec k, \omega_n) = - \left[ i\omega_n - \left(
\epsilon_{\vec k} +\sigma \Delta /2 -\mu^* \right) \right]^{-1}$.
The mean-field spin density is denoted by 
$n^{*}=(n_\downarrow - n_\uparrow )/2$, and 
$\epsilon_{\vec k} = \hbar^2 k^2/(2 m^*)$.

\begin{figure}
\centerline{\includegraphics[width=8cm]{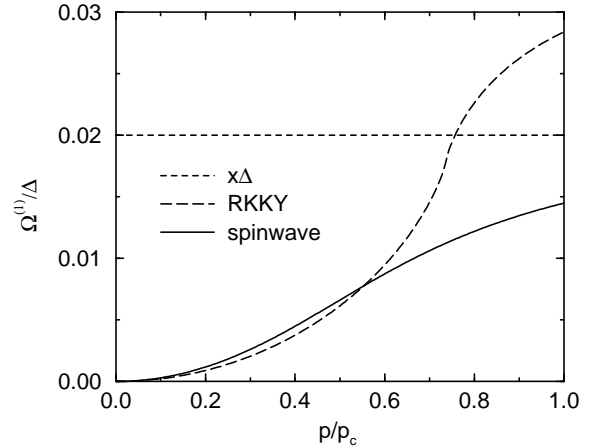}}
\caption{Spin-wave dispersion for $c^*= 0.1 {\rm nm}^{-3}$.
The short wavelength limit is the mean-field result $x\Delta$.
For comparison, we show also the result obtained from an RKKY picture.}
\label{fig1}
\end{figure}

{\it Excitations.}
We obtain the spectral density of the spin-fluctuation propagator by
analytical continuation, $i\nu_m \rightarrow \Omega+i0^+$ and 
$A(\vec p,\Omega)={\rm Im} \, D(\vec p,\Omega)/\pi$.
We find three different types of spin excitations.
In all figures we take $B=0$ and use as typical parameters\cite{Ohno99}
$m^*=0.5m_e$, $J_{pd}=0.15{\rm eVnm}^3$, and $c=1{\rm nm}^{-3}$, where $m_e$ is
the free-electron mass.
For these parameters the mean-field itinerant-carrier system is fully polarized
at $T=0$.

i) Our model has a gapless Goldstone-mode branch (see Fig.~\ref{fig1}) 
reflecting the spontaneous breaking of rotational symmetry\cite{gap}.
Expansion of the $T=0$ propagator at $\Delta > \epsilon_F$, where 
$\epsilon_F$ is the Fermi energy of the majority-spin band, yields for 
small and large momenta the dispersion of the collective modes
\begin{equation}
   \Omega^{(1)}_p = { x\over 1- x} \epsilon_p \left( 1 -
 {4 \epsilon_F\over 5\Delta} \right) + {\cal O}(p^4)
\end{equation}
and $\Omega^{(1)}_p = x \Delta ( 1 - \Delta / \epsilon_p ) + {\cal O} (1/p^4)$.
At short wavelengths we obtain the mean-field result $x\Delta$, the 
spin-splitting of a magnetic ion in the effective field produced by fully 
spin-polarized itinerant carriers.
Note that the itinerant-carrier and magnetic-ion mean-field spin splittings 
differ by a factor of $x=c^*/(2cS) \ll 1$.
At long wavelengths the magnon dispersion in an isotropic ferromagnet is 
proportional the spin stiffness $\rho$ divided by the magnetization $M$.
In the adiabatic limit\cite{Millis95,Nagaev98}, $\epsilon_F\ll\Delta$, our 
long-wavelength result reflects a spin-stiffness due entirely to the increase 
in kinetic energy of a fully spin-polarized band when the orientation has a 
spatial dependence, $\rho = c^* \hbar^2/(4m^*)$, and a magnetization $M$ which 
has opposing contributions from magnetic ions and itinerant carriers,
$M = c S -c^*/2 = c S (1-x)$.
In this limit, the mean-field critical temperature and the spin 
stiffness have opposite dependences on the itinerant-carrier mass.

\begin{figure}
\centerline{\includegraphics[width=8cm]{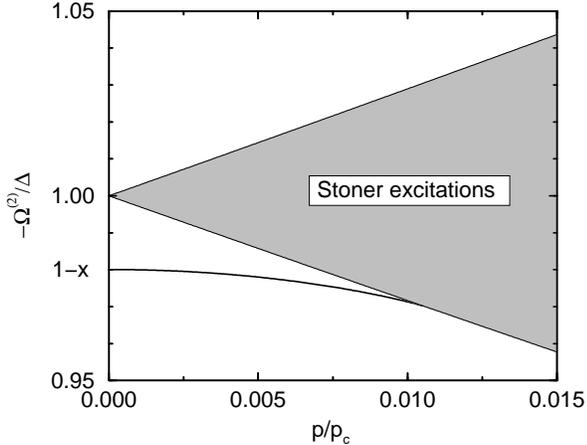}}
\caption{Stoner excitations and collective modes in the free-carrier system 
        for $c^*=0.1 {\rm nm}^{-3}$.}
\label{fig2}
\end{figure}

The low-energy mode describes spin waves in the local-impurity system.
This contrasts with the case of ferromagnetism induced by local 
moments\cite{Doniach67} where the low-energy mode is primarily in the 
free-carrier part.

ii) We find a continuum of Stoner spin-flip particle-hole excitations.
They correspond to flipping a single spin in the itinerant-carrier system and,
therefore, occur at much larger energies $\sim \Delta$ (see Fig.~\ref{fig2}).
For $\Delta > \epsilon_F$ and zero temperature, all these excitations carry 
spin $S^z=+1$, i.e., increase the spin polarization, and therefore turn up at
negative frequencies in the boson propagator we study.
(When $\Delta < \epsilon_F$, excitations with both $S^z=+1$ and $S^z=-1$ 
contribute to the spectral function.) 
This continuum lies between the curves $-\Delta-\epsilon_p\pm 
2\sqrt{\epsilon_p\epsilon_{F}}$ and for $\Delta < \epsilon_F$ also between
$-\Delta+\epsilon_p\pm 2\sqrt{\epsilon_p(\epsilon_F-\Delta)}$.

iii) We find additional collective modes associated primarily with the 
itinerant-carrier system at energies below the Stoner continuum 
(see Fig.~\ref{fig2}).
At $T=0$ we obtain
\begin{equation}
    -\Omega^{(2)}_p = \Delta (1-x) - { 1\over 1- x}
    \epsilon_p \left( {4 \epsilon_F\over 5x\Delta} -1 \right)
    + {\cal O}(p^4) 
\end{equation}
for $\Delta>\epsilon_F$.
The spectral weight of these modes is $-x/(1-x)$ at zero momentum.

The finite spectral weight at negative energies
indicates that the ground state is not fully spin polarized because of 
quantum fluctuations.

{\it Comparison to RKKY picture.} 
For comparison we evaluate the $T=0$ magnon dispersion assuming an RKKY 
interaction between magnetic ions.  
This approximation results from our theory if we neglect the spin polarization 
in the itinerant carriers and evaluate the static limit of the resulting 
spin-wave propagator.
The Stoner excitations and magnons shown in Fig.~\ref{fig2} do not emerge 
and the spin-wave dispersion (Fig.~\ref{fig1}) is incorrect except for the 
limit $\Delta\ll\epsilon_F$. 
While the RKKY picture does, in some circumstances, provide a realistic 
estimate of $T_c$, it completely fails as a theory of the ferromagnetic state.

{\it Comparison to a ferrimagnet.}
Some features of the excitation spectrum are, not coincidentally, like those 
of a localized spin ferrimagnet with antiferromagnetically coupled large spin 
$S$ (corresponding to the magnetic ions) and small spin $s$ (corresponding to 
the free carriers) subsystems on a bipartite cubic lattice.
Representing the spins by HP bosons and expanding up to quadratic order,
we either diagonalize the resulting Hamiltonian directly or, as in our 
ferromagnetic-semiconductor calculations, integrate out the smaller spins 
using a path-integral formulation to obtain equivalent results.  
We find that
\begin{equation}
   \Omega^{(1)/(2)}_{\vec p} = {\Delta\over 2} \left[ -(1-x) \pm
 \sqrt{(1-x)^2+4x\gamma_{\vec p}} \, \right]
\end{equation}
with $x=s/S$ and $\Delta=6JS$, where $J$ is the exchange coupling, and 
$\gamma_{\vec p}=(1/3)\sum_{i}[1-\cos(p_{i}a)]$ with lattice constant $a$.
We recover two collective modes, the coupled spin waves of the two subsystems.
One is gapless, the other one gapped with $\Delta(1-x)$.
The bandwidth is $x\Delta$, and the spectral weights at zero momentum are 
$1/(1-x)$ and $-x/(1-x)$, respectively, as in our model.

{\it Self-consistent scheme.}
Near the transition temperature $T_c$, the spin-wave density is of the 
order of $cS$, and expanding around a fully-polarized state is not a good 
starting point.
Instead, we adopt a scheme which accounts self-consistently for the reduction 
of spin density in the localized impurity and itinerant carrier 
subsystems\cite{foot}.
Our approach is motivated by the Weiss mean-field theory, which is expected to 
be accurate for a model with static long-range interactions between the 
magnetic ions.
This model can be obtained in our approach by taking the Ising limit (i.e., 
replacing $\vec S \cdot \vec s$ by $S^z s^z$) and letting the mean-field 
itinerant-carrier spin splitting, $\Delta(T)= J_{pd} \langle S^z \rangle$,
reflect the thermal suppression of the impurity-spin 
density $\langle S^z \rangle$.
In this simple model the spin-wave spectrum is independent of momentum, 
$\Omega_{p}(T) = J_{pd} n^{*}(T)$, allowing a unitary transformation to 
spin-wave eigenstates with a site label.
The constraint on the number of spin bosons ($ \leq 2S$) on a site is then 
easily applied and we obtain 
\begin{eqnarray}
   &&\langle S^z \rangle = {1\over V}\sum_{|\vec p| < p_c}
 S B_S (\beta S \Omega_p)
\nonumber\\
   && = {1\over V}\sum_{|\vec p| < p_c} \left\{ S - n(\Omega_p)
 + (2S+1) n[(2S+1)\Omega_p] \right\},
\label{SelfConsistent}
\end{eqnarray}
where $B_S(x)$ is the Brillouin function, and $n(\omega)$ is the Bose function.
The first two terms in the second form of Eq.~(\ref{SelfConsistent}) give the 
spin density obtained by treating spin waves as bosons, while the last term is 
the correction from spin kinematics which rules out unphysical states.

\begin{figure}
\centerline{\includegraphics[width=7.9cm]{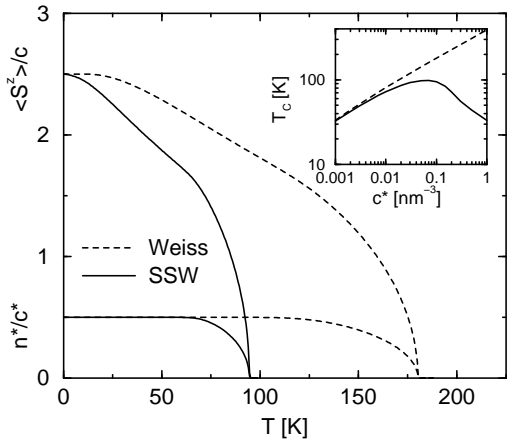}}
\caption{Average impurity and itinerant-carrier spin as a function of 
        temperature for $c^*=0.1{\rm nm}^{-3}$ in Weiss mean-field (Weiss) and 
        self-consistent spin-wave theory (SSW). 
        Inset: $T_c$ as a function of $c^*$.}
\label{fig3}
\end{figure}

Our self-consistent spin-wave approximation consists of using 
Eq.~(\ref{SelfConsistent}) when $\Omega_p$ is $p-$dependent 
(see Fig.~\ref{fig3}).
Although the characteristic $T^{3/2}$ law for the localized-ions magnetization 
is recovered at low temperatures, the prefactor is reduced by 
$1-(2S+1)^{-1/2}$, compared to the correct\cite{Dyson56} value of linearized 
spin-wave theory.
That is, constraining boson populations in momentum space is too restrictive 
at low temperatures, although the error is not serious for $S=5/2$.
The transition temperature of the self-consistent spin-wave theory is lower 
than the estimate from Weiss mean-field theory.

\begin{figure}
\centerline{\includegraphics[width=7.8cm]{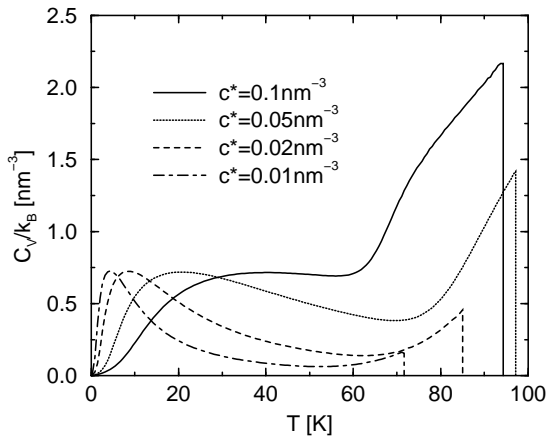}}
\caption{Specific heat due to spin waves.}
\label{fig4}
\end{figure}

The presence of spin waves also shows up in the specific heat $C_V$, which we 
calculate within the same scheme (see Fig.~\ref{fig4}).
Calculating the entropy $S$ we obtain
\begin{equation}
    C_V = {T\over V}{dS\over dT} =
         \frac{1}{V} \sum_{|\vec p|<p_{c}} \Omega_{p}(T)
    \left(\frac{dN_{p}}{dT} \right),
\end{equation}
where $N_{p} = S - S B_{S}(\beta S \Omega_{p})$ is the average number
of spin bosons at each momentum. 
The specific heat from the kinetic energy of itinerant carriers turns out to 
be negligibly small. 
The specific heat of magnetic ions $C_{V}$ is proportional to $T^{3/2}$ at low
temperature, shows a Schottky-like anomaly, and has a jump at the transition 
temperature.
We estimate that the jump is 5\% of the lattice specific heat 
when $T_c \sim 100{\rm K}$, suggesting that this should be observable.

We acknowledge useful discussions with M. Abolfath, B. Beschoten, A. Burkov, 
J. Furdyna, S. Girvin, and T. Jungwirth.
This work was supported by the Deutsche Forschungsgemeinschaft under grant
KO 1987-1/1 and by the National Science Foundation, DMR-9714055.


\begin{thebibliography}{10}
\bibitem{foot}
In the following, we neglect Stoner excitations and magnons in the itinerant 
carriers, i.e., we assume sharp spin-wave excitations with spectral weight one.
\end{thebibliography}
\end{document}